\documentstyle[aps,preprint,epsfig,float]{revtex}

\begin{document}
\title{Hydrodynamic Instability of the Flux-antiflux Interface in Type-II
Superconductors}
\author{L.M. Fisher}
\address{All-Russian Electrical Engineering Institute,
12 Krasnokazarmennaya Street, 111250 Moscow, Russian Federation}
\author{P.E. Goa, M. Baziljevich, T.H. Johansen}
\address{Department of Physics, University of Oslo, P.O. Box 1048, Blindern, 0316
Oslo 3, Norway}
\author{A.L. Rakhmanov}
\address{Institute for Theoretical and Applied Electrodynamics Russian Academy
of Science, 13/19 Izhorskaya Street, 127412 Moscow, Russian Federation}
\author{V.A. Yampol'skii}
\address{Institute for Radiophysics and Electronics Ukrainian Academy of
Science,12 Proskura Street, 61085 Kharkov, Ukraine}

\maketitle

\begin{abstract}
The macroturbulence instability observed in fluxline systems during remagnetization
of superconductors is explained. It is shown that when a region with flux is invaded
by antiflux the interface can become unstable
if there is a relative tangential flux motion. This condition occurs at the interface
when the viscosity is anisotropic, e.g., due to flux guiding by twin boundaries
in crystals. The phenomenon is similar to the instability of the tangential
discontinuity in classical hydrodynamics. The obtained results are supported by
magneto-optical observations of flux distribution on the surface of a YBCO single
crystal with twins.
\end{abstract}
\pacs{74.25.Ha, 74.60.Ge, 74.60.Jg}

The interest in the
physics of the vortex state in type-II superconductors increased
significantly after the discovery of high-$T_c$ superconductivity
(HTS). A main reason for the renewed attention is the observation
of many novel nontrivial phenomena occurring in the vortex matter
of the HTS materials. The perhaps most dramatic of these phenomena
is the turbulence instability of the vortex-antivortex interface,
which was observed in 1--2--3 systems using magneto-optical (MO)
imaging~\cite{vl,ind,joh}. It consists in the following. When
magnetic flux is trapped in the superconductor and a moderate
field of reverse direction is subsequently applied, a boundary of
zero flux density will separate regions containing flux and
antiflux. In some temperature and field range this flux-antiflux
distribution can display unstable behavior characterized by an
irregular propagation of the boundary where finger-like patterns
develop. This contrasts strongly the regular propagation of the
flux front during virgin field penetration, when only one flux
polarity is present in the sample. It is clear that the
instability, often called macroturbulence in the literature,
cannot be understood within the frameworks of the critical state
model or conventional models for flux relaxation~\cite{bean,yes}.

An attempt to explain this remarkable behavior of the
flux-antiflux interface was made in~\cite{bass}, where the
instability was attributed to a thermal wave generated by local
heat release in the vortex-antivortex annihilation. Unfortunately,
this mechanism can hardly be accepted. Indeed, the vortex energy
consists of two terms, where one is the magnetic part related to
the magnetic field of the vortices. The other part is stored
within the vortex core and represents the condensation energy. The
magnetic energy of a vortex-antivortex pair is dissipated as Joule
heat as they are getting close to each other, but before the
annihilation takes place. This energy, which equals ${\vec J}{\vec
E}$ where ${\vec J}$ is the current density and ${\vec E}$ is the
electric field generated by the vortex motion, is dissipates not
only at the interface, but in the bulk of the sample. Hence, only
the release of the core energy in the process of annihilation is
concentrated near the interface. A simple calculation shows that
the core energy is much smaller than the magnetic part and will
cause only a negligible rise of the sample
temperature~\cite{comment}, thereby ruling out that thermal
effects are responsible for the instability.

In this Letter we present an explanation focussing on the
experimental fact that the instability was reported only for ${\rm
YBa_2Cu_3O_{7-\delta}}$ and other 1--2--3 single crystals. These
materials are characterized by the existence of two twin boundary
systems oriented orthogonally to each other in the {\bf ab} plane.
The vortex motion in such crystals is governed by a pronounced
guiding effect~\cite{guid1,guid2}, and the viscosity coefficient
for motion along the twin boundaries should be much smaller than
across them. This anisotropy gives rise to vortex motion with a
velocity component normal to the Lorentz force. The vortices and
antivortices are then forced to move towards each other along the
interface, where the tangential component of the velocity becomes
discontinuous. It is well known that hydrodynamical flow under
such conditions can be unstable~\cite{lan}, and we show here that
a purely hydrodynamic theory of the vortex-antivortex system with
anisotropic viscosity can explain the origin of the
macroturbulence.

Consider an infinite superconducting plate of thickness
$2d$ with the external magnetic field ${\vec H}$ oriented parallel
to the sample surface along the $z$-axis. The $x$-axis is
perpendicular to the plate and $x=0$ in the center.
Let $H$ first increase and then be lowered through zero to
a negative value. Then two kinds of vortices will exist in the
sample; one with field direction along the positive $z$-axis
({\it vortices}) and one directed oppositely ({\it antivortices}).
From the symmetry of the problem it is sufficient to consider
only the region $0<x<d$, and Fig.~1
%Fig.~\ref{f1}
shows schematically the distributions $N_1(x)$ and $N_2(x)$ of
vortex and antivortex densities.

The densities $N_1$ and $N_2$ satisfy the continuity equation,
\begin{equation}  \label{1}
\frac{\partial N_\alpha}{\partial t} + {\rm div}(N_\alpha{\vec V}_\alpha)=0,
\qquad \alpha = 1, 2
\end{equation}
where ${\vec V}_\alpha$ are the vortex and antivortex velocities. In the
regime of anisotropic viscous flow they are related to the Lorentz
driving force by,
\begin{equation}  \label{2}
\eta_{ik}N_\alpha V_{\alpha k}=F_{L i}, \qquad {\vec F}_L =
\frac{1}{c}{\vec B}\times {\vec J},
\end{equation}
\vspace*{-4mm}
\[
{\vec B}=N_\alpha {\vec \Phi}_0, \qquad {\vec J}=
\frac{c}{4\pi}\nabla\times{\vec B}.
\]
Here $\eta_{ik}$ is the symmetrical tensor of the anisotropic viscosity
and $\Phi_0$ is the magnetic flux quantum. Eq.~(\ref{2}) can be rewritten as,
\begin{equation}  \label{3}
V_{\alpha i}=-\frac{\Gamma\Phi_0^2}{4\pi}\gamma_{ik}
\frac{\partial N_\alpha}{\partial x_k},
\end{equation}
where $\gamma_{ik}$ is the dimensionless tensor of the inverse
viscosity, $\eta_{ik}^{-1}=\Gamma\gamma_{ik}$, and the coefficient $\Gamma$
is chosen so that the principal values of $\gamma_{ik}$ are unity
and $\varepsilon$, which  satisfies $0<\varepsilon<1$.
The case of a strong guiding effect
corresponds to $\varepsilon\to 0$. Note that $\Gamma$ increases rapidly
with the temperature due to thermal depinning of the vortices.

To solve the problem one must also formulate boundary conditions at the
vortex-antivortex interface. Generally the position of the interface, $x=x_0$,
depends on the $y$-coordinate and time $t$ as it moves with a
velocity ${\vec U}$.
The first condition is that
the total flux of vortices and antivortices through
the interface vanishes,
\begin{equation}  \label{4}
N_1({\vec V}_1 -{\vec U})_n + N_2({\vec V}_2 -{\vec U})_n =0.
\end{equation}
Secondly, both vortex polarities annihilate at the interface with the rate
proportional to the product of their densities,
\begin{equation}  \label{5}
N_1({\vec V}_1 -{\vec U})_n = RN_1N_2.
\end{equation}
The parameter $R$ can depend on the vortex densities and velocities but, for
simplicity, we assume here that it is a phenomenological constant. A similar
model for the annihilation process was used in Ref.\onlinecite{bass}. However,
contrary to~\cite{bass}, we consider the parameter $R$ to be defined by
the microscopic Meissner current of individual vortices,
which is much greater than the macroscopic currents $J$. Therefore, the
relative velocity of annihilating vortices and antivortices is much greater
than the hydrodynamic velocity $V_\alpha$.
As a result, the region where vortices and
antivortices coexist and annihilation takes place is very narrow, and
can be represented by the surface $x=x_0(y,t)$.
Finally, we take the magnetic induction to be zero, i. e.,
\begin{equation}  \label{6}
N_1=N_2 ,
\end{equation}
at $x=x_0 (y,t)$. This condition is also a direct consequence of
Eq.~(\ref{4}) if the vortex and antivortex densities on the interface
were equal to zero initially.

As a first step, we determine what the model predicts for the
unperturbed distribution profiles $N_1$ and $N_2$.
In this case the vortex-antivortex interface is the plane
$x=x_0(t)$ which moves with the velocity $U={\rm d}x_0(t)/{\rm
d}t$. Since a simple solution
with constant $U\ne 0$ could not be found, we consider only
the stationary profile with $U=0$. With
$\partial N_1/\partial t= \partial N_2/\partial t=0$,
the Eqs.~(\ref{1}), (\ref{4})--(\ref {6}) give,
$$
N_1(x)=N_2(d)\left(\frac{x_0+d/2r-x}{d+d/2r-x_0}\right)^{1/2},
$$
\begin{equation}  \label{7a}
N_2(d)=H_0/ \Phi_0,
\end{equation}
\vspace*{-4mm}
$$
N_2(x)=N_2(d)\left(\frac{x+d/2r-x_0}{d+d/2r-x_0}\right)^{1/2},\quad
r=\frac{4\pi Rd}{\Gamma\Phi_0^2\gamma_{xx}}.
$$
In the following, we
assume $r\gg 1$ since the rate of the annihilation is fast and the viscosity
is not small. Therefore, at the interface the vortex densities are
relatively small, $N_1=N_2 \sim N_2(d)r^{-1/2}$ while the spatial derivatives
of $N_1$ and $N_2$ at $x=x_0$ are large,
\begin{equation}  \label{9}
N_2^\prime= -N_1^\prime=N_1r/d, \qquad N_2^{\prime\prime}=
N_1^{\prime\prime}=-N_1(r/d)^2.
\end{equation}
Then, also the velocities $V_{\alpha\; x,y}(x=x_0)$ are relatively high,
since they are proportional to $r^{1/2}$ near the interface.

If one cannot provide fixed values of the vortex and antivortex
densities $N_1(0)$ and $N_2(d)$ at the boundaries,
the discussed stationary profiles cannot be realized. In practice the
external magnetic field defines the density of the antivortices only.
Because of this asymmetry and the vortex-antivortex annihilation, the total number
of the vortices will decrease with time and the interface moves towards the middle of
the sample. To simplify the problem, we assume the interface velocity $U$ to
be much smaller than the vortex velocity $V_\alpha$, i.\ e.\ we consider the
problem in the quasi-stationary regime.

To investigate the stability of the interface with respect to
small perturbations, it is suitable to introduce the following dimensionless
variables,
\begin{eqnarray}  \label{11}
\displaystyle n_\alpha& =& N_\alpha/N_\alpha(x_0), \quad
\tau=t/t_0, \quad t_0= \frac{\Gamma \Phi_0^2 \gamma_{xx}}{4\pi R^2
N_\alpha(x_0)},  \nonumber\\ \noindent \xi&=&x/L, \quad \zeta=y/L,
\quad L=\frac{\Gamma \Phi_0^2 \gamma_{xx}}{4\pi R}=d/r.
\end{eqnarray}
Normalization using the time-dependent $N_\alpha(x_0(t))$ is
allowed here since we assume that the instability develops much
faster than noticeable changes occur in $N_\alpha(x_0)$. We seek
for perturbations in the vortex and antivortex densities of the
form
\begin{equation}  \label{12}
n_\alpha = n_\alpha^{(0)} + f_\alpha \exp [\lambda \tau +{\rm i} k \zeta +
p_\alpha(\xi-\xi_0(\tau))].
\end{equation}
The linearized boundary conditions should be written on the
perturbed interface,
\begin{equation}  \label{13}
\xi=\xi_0(\zeta, \tau)=\xi_0(\tau)+ \delta \xi \exp({\rm i}k\zeta
+\lambda \tau),
\end{equation}
with the normal unit vector
\begin{equation}  \label{14}
{\vec \nu}=(1, \; -{\rm i} k\delta \xi(\zeta,\tau)).
\end{equation}
It follows directly from Eq.~(\ref{6}) that
\begin{equation}  \label{15}
\delta \xi = (f_1 - f_2)/2.
\end{equation}

Equations~(\ref{1}) give the expressions for the parameters $p_1$ and $p_2$.
Substituting them and Eqs.~(\ref{12}), (\ref{15}) into Eqs.~(\ref{4}), (\ref
{5}), we obtain two linear algebraic homogeneous equations relating
$f_1$ and $f_2$. Demanding the determinant to vanish and omitting
 terms of the order of $(U t_0/L)^2\ll 1$, one obtains the following dispersion
equation for the increment $\lambda$ at different wave numbers $k$,
\begin{equation}  \label{16}
\lambda=\Omega^2-\epsilon \kappa^2 -2{\rm i}s\kappa -1 -b \ .
\end{equation}
Here
\begin{equation}  \label{17}
\kappa = \frac{k|\alpha u|}{2}, \quad \alpha =
\frac{\gamma_{xy}}{\gamma_{xx}}, \quad u=U \frac{t_0}{L},
\end{equation}
\[
\epsilon = 4\varepsilon/(\alpha u)^2, \quad s = {\rm sign}(\alpha u), \quad
2b=-n_1^{\prime \prime}- n_2^{\prime \prime},
\]
and $\Omega$ is a root with ${\rm Re}\Omega >0$ of the equation
\begin{eqnarray}  \label{18}
\displaystyle \Omega^4 + 3\Omega^3 +\Omega^2(-\epsilon \kappa^2 +2
-2b)  \nonumber \\ -\Omega (2\epsilon \kappa^2 +{\rm i}s\kappa
+4b) -3s{\rm i}\kappa=0.
\end{eqnarray}

Shown in Fig.~2
%Fig.~\ref{f2}
is the dependence of the increment ${\rm Re}\lambda$ on the
dimensionless wave number $\kappa$ for different $\epsilon$. The
parameter $b$ is set to unity as this follows from the
quasi-stationarity condition $u\to 0$.
The curves demonstrate that a positive increment exists, i.e. the planar
interface becomes unstable, when $\epsilon
<\epsilon _{c}= 0.019$. The instability occurs at not too small values of
the anisotropy and velocity of the flux-antiflux interface. Furthermore, it
is characterised by a temporal scale given by the
largest ${\rm Re}\lambda$, $\lambda_{m}$, which
occurs at finite $\kappa =\kappa _{m}$. For $\epsilon$ sufficiently small,
$\kappa _{m}$ is much greater than unity and one finds
from Eq.~(\ref{18})
\begin{equation}
\lambda_{m}=1/(4\sqrt{\epsilon})-2,\qquad
\kappa_{m}=1/\sqrt{2}\epsilon ^{3/4}.  \label{19}
\end{equation}

Thus, we have identified that an instability occurs if
$\epsilon <\epsilon_{c}$,
which in dimensional notation is expressed as
\begin{equation}
\varepsilon <\epsilon_{c}\left[\frac{U
\tan\theta}{2RN_{1}(x_{0})}\right] ^{2} \label{20}
\end{equation}
where $\theta$ is the angle between the direction of flux guiding and
the flux-antiflux interface. The above analysis allows us to understand
why the instability of the vortex-antivortex interface has been
observed only in crystals of the 1--2--3 system, where a pronounced
guiding effect is expected due to the twin boundaries.

As an illustration we show in Fig.~3
%Fig.~\ref{f3}
turbulent behaviour observed in an optimally doped YBCO single
crystal containing a substantial amount of twinning, see (a). The
crystal has a rectangular shape in the {\bf ab} plane and measures
1~mm along the longest edge.
Shown in (b) is the MO-image of flux penetration
in an external field of 100 mT applied after zero-field-cooling to
45 K. One sees that the field penetrates predominantly from some
large twin boundaries located at the bright core of the lines that
make $45^{\circ}$-angles with the edges of the crystal. The dark
area in the centre of the sample is the region not being
penetrated by a 100~mT applied field at this temperature. In (c)
the temperature is raised to 67 K, and the MO-image was recorded
in the remanent state after first applying 100~mT. The bright
"aura" around the crystal is here the return field of the flux
trapped in the central part. Note that this reverse field partly
penetrates the sample near the edge. A distinct line can be seen
as a dark band going around the crystal just on the inside of the
edge. This band is the annihilation zone, which divides the
crystal into two opposite magnetic domains. The macroturbulence is
here seen as a meandering of the annihilation zone.
Adjacent to the zone one can find small areas with increased flux
density, see magnified view in (d). Time resolved
measurements show that the zone develops in a highly dynamic
manner where abrupt redistributions of flux often occur.
By adding an external reverse field to the remanent state the
annihilation zone is pushed further into the
crystal and the dynamical features becomes even more spectacular.

In this crystal the turbulent behavior was observed in the interval
25-75~K. As the temperature increases the dynamics of the
flux/antiflux interface becomes increasingly rapid. However, above
75~K there again appears to be no irregular behavior of the
interface.

The existence of the instability only in a definite temperature
region finds a simple rationalization within our model.
At low temperatures, the viscosity increases exponentially, and
the characteristic spatial scale $L$, in Eq.~(\ref{11}), decreases
correspondingly and becomes comparable to or less than the
twin-boundary spacing. As a result, the guiding effect is
suppressed and the instability disappears. On the other hand,
at temperatures close to $T_{c}$ the flux guiding is no longer
effective due to thermal activation of the vortices.
It is remarkable, and in full support of our model,
that in the present heavily twinned crystal
the turbulence occurs down to much lower temperatures than found in
previous studies of similar crystals with only little
twinning~\cite{koblisch,joh}.

In conclusion, we have presented a theoretical model which gives
a fully consistent description of the macroturbulence phenomenon
we observe in HTS materials under certain conditions.
This work is supported by INTAS/RFBR and RFBR, grants IR-97--1394
and 00-02-17145, 00-02-18032, and the Research Council of Norway.

\newpage
%\section*{Figures}\vspace{1cm}
\begin{figure}[t]
\unitlength1cm
\begin{center}
\begin{picture}(8,9)
\epsfxsize=8cm \put(0,1){\epsfbox{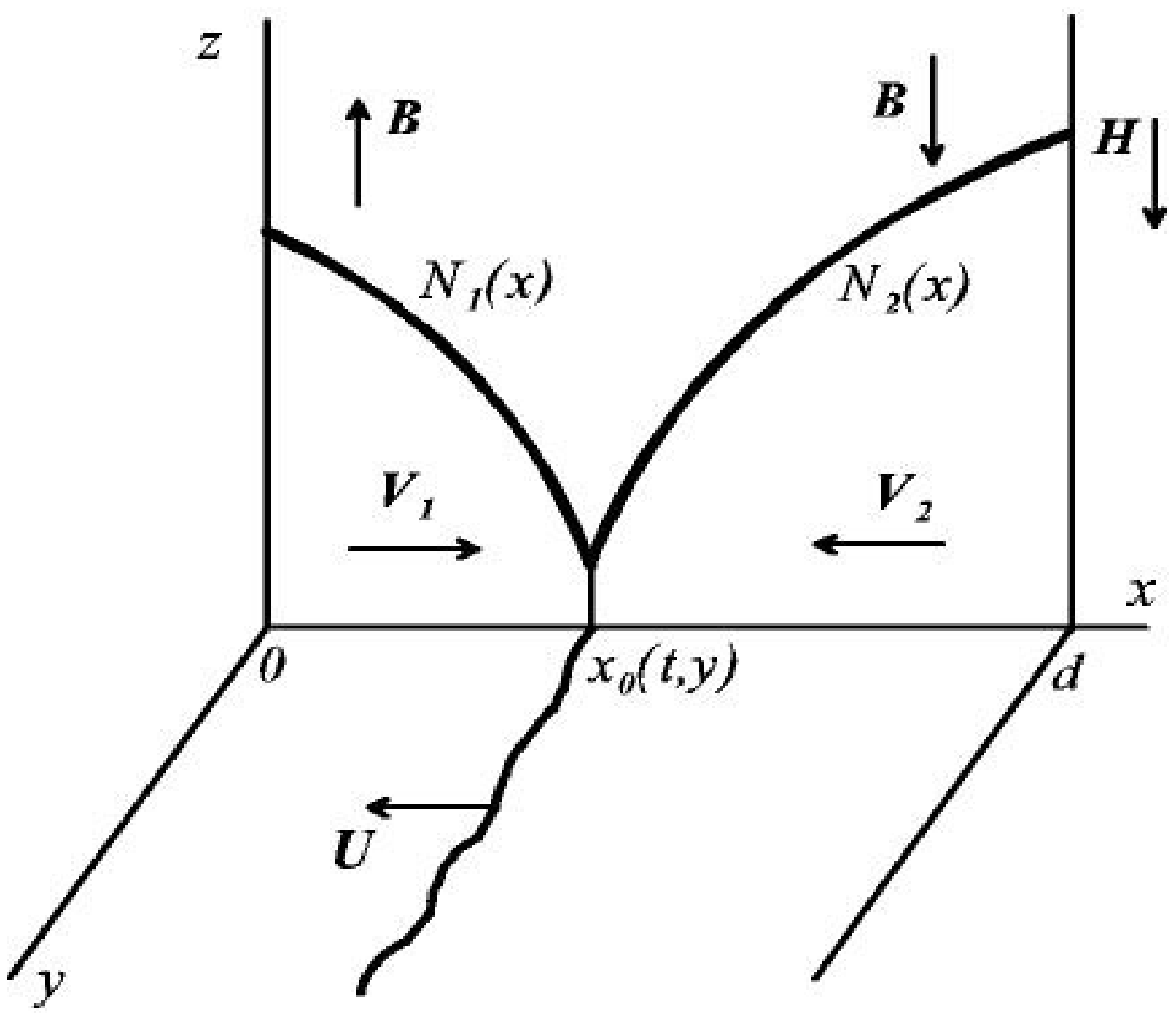}}
\end{picture}
\caption[]{\label{model}Flux distribution in one half of an
infinite slab $(|x|\leq d)$ containing trapped vortices of density
$N_1(x)$ in the central region $|x| \leq x_0$, and antivortices of
density $N_2(x)$ penetrating from the outside. The other symbols
are defined in the text. }
\end{center}
\end{figure}

%\newpage
\begin{figure}[t]
\unitlength1cm
\begin{center}
\begin{picture}(8,8)
\epsfxsize=10cm \put(-0.5,-1.5){\epsfbox{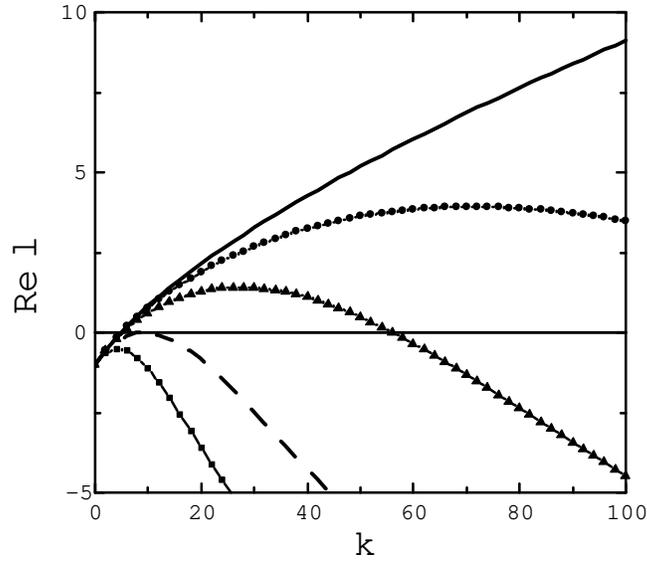}}
\end{picture}
\end{center}
\caption[]{\label{model}The dependence $\mathrm{Re}\lambda
(\kappa)$ at different anisotropy parameters $\varepsilon$:
$\varepsilon = 0$ (solid line), $\varepsilon = 0.0015$ (circles),
$\varepsilon = 0.005$ (triangles), $\varepsilon = \varepsilon_c =
0.019$ (dashed line), $\varepsilon = 0.05$ (boxes). }
\end{figure}

\newpage
\begin{figure}[t]
\unitlength1cm
%\begin{center}
\begin{picture}(8,8)
\epsfxsize=15cm \put(0,2){\epsfbox{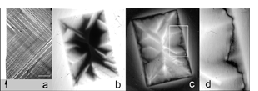}}
\end{picture}
%\end{center}
\caption[]{\label{model}(a) - Polarized light image showing twin
domains in a small area on the crystal. The arrow indicates the
sample edge, and the scale bar is 50~$\mu$m long. (b)-(d) are
magneto-optical images where the brightness represents the
magnitude of ${\bf B}$'s component normal to the surface. (b) -
Applied field of $B_a = 100$ mT at $T= 45$ K. (c) - Remanent state
after full flux penetration at $T = 67$ K. (d) - magnified view of
the area marked in (c) }
\end{figure}


\begin{thebibliography}{99}
\bibitem{vl}  V.K.~Vlasko-Vlasov, V.I.~Nikitenko, A.A.~Polyanskii,
G.W.~Grabtree, U.~Welp, and B.W.~Veal, Physica C {\bf 222}, 361 (1994).

\bibitem{ind}  M.V.~In\-den\-bom, Th. Schuster, M.R.~Kob\-lisch\-ka, A.~Forkl,
H.~Kronm\"{u}ller, L.A.~Do\-ro\-sin\-skii, V.K.~Vlas\-ko-Vla\-sov,
A.A.~Po\-lyan\-skii, R.L.~Pro\-zo\-rov, and V.I.~Ni\-ki\-ten\-ko,
Physica C {\bf 209}, 259 (1993).

\bibitem{joh}  T.~Frello, M.~Baziljevich, T.H.~Johansen, N.H.~Andersen,
Th.~Wolf, and M.R.~Kob\-lisch\-ka, \prb {\bf 59}, R6639 (1999).

\bibitem{bean}  C.P.~Bean, \prl {\bf 8}, 250 (1962).

\bibitem{yes}  Y.~Yeshurun, A.P.~Malozemoff, and A.~Shaulov, Rev. Mod. Phys.
{\bf 68}, 911 (1996).

\bibitem{bass}  F.~Bass, B.Ya.~Shapiro, I.~Shapiro, and M.~Shvartser,
\prb {\bf 58}, 2878 (1998).

\bibitem{comment}
The ratio of the core energy and magnetic energy of a vortex is
$W_c/W_J \approx H_{c1}^2/H_{c2}H_0\ll 1$ ($H_{c1}$ and $H_{c2}$
are the lower and upper critical magnetic fields, and $H_0$ is the
magnetic field on the sample surface). Release of the core energy
of all vortices in the sample would increase the temperature only
on the order of $10^{-4}$~K.

\bibitem{guid1}A.K. Niessen, C.H.~Weijsenfeld, J. Appl. Phys., {\bf
40}, 384 (1969).
\bibitem{guid2} H. Pastoriza, S Candia, G. Nieva, \prl, {\bf 83}, 1026
(1999).
\bibitem{lan}  L.D. Landau and E.M. Lifshits, {\em Fluid Mechanics}
(Butterworth-Heinemann, Oxford, 1987).

\bibitem{koblisch}  M.R. Koblischka, T.H. Johansen, M. Baziljevich,
H. Hauglin, H. Bratsberg and B.Ya. Shapiro, Europhys. Lett. {\bf 41},
419 (1998).

\end{thebibliography}
\end{document}